\documentclass{emulateapj}
\usepackage{epsfig}
\usepackage{subfigure}
\usepackage{amssymb}
\usepackage{color}
\slugcomment{To be published by the Astrophysical Journal Letters}

\def\gtorder{\mathrel{\raise.3ex\hbox{$>$}\mkern-14mu
    \lower0.6ex\hbox{$\sim$}}}
\def\ltorder{\mathrel{\raise.3ex\hbox{$<$}\mkern-14mu
    \lower0.6ex\hbox{$\sim$}}}

\shorttitle{Damping Stellar Bars In Spinning Halos}
\shortauthors{Long, Shlosman and Heller}

\begin{document} 

\title{Secular Damping of Stellar Bars In Spinning Dark Matter Halos}

\author{ 
Stacy Long\altaffilmark{1}, 
Isaac Shlosman\altaffilmark{1,2}, 
Clayton Heller\altaffilmark{3}
}
\altaffiltext{1}{
Department of Physics and Astronomy, 
University of Kentucky, 
Lexington, KY 40506-0055, 
USA
}
\altaffiltext{2}{
Theoretical Astrophysics,
Department of Earth and Space Science,
Osaka University,
Osaka 560-0043,
Japan
}
\altaffiltext{3}{
Department of Physics, 
Georgia Southern University, 
Statesboro, GA 30460, 
USA
}

\begin{abstract}
We demonstrate that growth of stellar bars in spinning dark matter halos is heavily
suppressed in the secular phase of evolution, using numerical simulations of
isolated galaxies. In a representative set of models, we show that for values of the
cosmological spin parameter $\lambda\gtorder 0.03$, bar growth (in strength and size)
becomes increasingly quenched. Furthermore, slowdown of bar pattern speed weakens
substantially with increasing $\lambda$, until it ceases completely. The terminal
structure of the bars is affected as well, including extent and shape of their boxy/peanut
bulges. The essence of this effect lies in the modified angular momentum exchange
between the disk and the halo facilitated by the bar. For the first time we have
demonstrated that a dark matter halo can emit and not purely absorb angular momentum.
Although the halo as a {\em whole} is not found to emit, the net transfer of angular
momentum from the disk to the halo is significantly reduced or completely eliminated.
The paradigm shift implies that the accepted view that disks serve as
sources of angular momentum and halos serve as sinks, must be revised. Halos with
$\lambda\gtorder 0.03$ are expected to form a substantial fraction, based on lognormal
distribution of $\lambda$. Dependence of secular bar evolution on halo spin,
therefore, implies profound corollaries for the cosmological evolution of galactic
disks.
\end{abstract}

\keywords{dark matter ---galaxies: evolution --- galaxies: formation --- galaxies: halos --- 
galaxies: interactions --- galaxies: kinematics and dynamics}
    
\section{Introduction}
\label{sec:intro}

Redistribution of angular momentum in astrophysical systems is a major driver of their dynamical and
secular evolution. Galactic bars facilitate this process  by means of gravitational torques, triggered 
internally (spontaneously) or externally (interactively). Important aspects of stellar bar evolution
are still being debated --- their origin and evolutionary changes in 
morphology, growth and decay, are not entirely clear. 
Theoretical studies of angular momentum redistribution in disk-halo systems
have been limited almost exclusively to {\it nonrotating} halos, following pioneering 
works on linear perturbation theory by Lynden-Bell (1962), Lynden-Bell \& Kalnajs (1972), Tremaine \& 
Weinberg (1984) and Weinberg (1985), which underscored the dominant role of orbital resonances. Numerical 
simulations have confirmed the angular momentum flow away from disks embedded in
axisymmetric (e.g., Sellwood 1980; Debattista \& Sellwood 1998, 2000; Tremaine \& Ostriker 1999;
Villa-Vargas et al. 2009, 2010; review by Shlosman 2013) and triaxial 
(e.g., El-Zant \& Shlosman 2002; El-Zant et al. 2003; Berentzen et al. 2006; Berentzen 
\& Shlosman 2006; Heller et al. 2007; Machado \& Athanassoula 2010; Athanassoula et al. 
2013) halos. Resonances have been confirmed to account for the lion's share of angular momentum 
transfer (e.g., Athanassoula 2002, 2003; Martinez-Valpuesta et al. 2006; Weinberg \& Katz 2007), 
In this paradigm, the halo serves as the pure sink and the disk as the net source of angular  momentum.

However, realistic cosmological halos are expected to possess a net angular momentum, acquired 
during the maximum expansion epoch (e.g., Hoyle 1949; White 1978) and possibly during
the subsequent evolution (Barnes \& Efstathiou 1987; but see Porciani et al. 2002). Simulations 
have quantified the distribution of spin values,
$\lambda\equiv J_{\rm h}/\sqrt{2} M_{\rm vir}R_{\rm vir}v_{\rm c}$, for cosmological dark matter (DM) 
halos to follow a lognormal distribution, where $J_{\rm h}$ is the
angular momentum, $M_{\rm vir}$ and $R_{\rm vir}$ --- the halo virial mass and radius, and $v_{\rm c}$
--- the circular velocity at $R_{\rm vir}$,  with the mean value $\lambda = 0.035\pm 0.005$ (e.g., 
Bullock et
al. 2001). Spinning halos can increase the rate of the angular momentum absorption --- 
an issue brought up by Weinberg (1985) but never fully addressed since. Only recently has it been
confirmed numerically that the bar instability timescale is indeed shortened for $\lambda>0$
(Saha \& Naab 2013). But these models had been terminated immediately after the bar instability had 
reached its peak, and hence avoided completely the secular stage of bar evolution.

The $\lambda=0$ halos consist of two populations of DM particles, prograde and retrograde
(with respect to disk spin). The amount of angular momentum in {\it each} of these populations can vary
from zero for nearly radial orbits, to a maximal one for nearly circular orbits. (Both extremes are 
mentioned 
for pedagogical reasons only.) These extremes in angular momentum correspond to extremes in velocity
anisotropy.  Various degrees of velocity anisotropy 
in the halo lie in between, and represent a rich variety of dynamical models. 
Stellar bars mediate the angular momentum transfer in such disk-halo systems with a broad range
of efficiencies. The current paradigm of stellar bar evolution assumes an idealized 
version of a nonrotating DM halo which cannot account for the whole bounty of associated processes. 
We address these issues in a subsequent paper (in preparation).

In this Letter we demonstrate for the first time that secular growth of galactic bars in spinning DM 
halos is damped more strongly with increasing $\lambda$, and this effect is the result of a modified 
angular momentum transfer. Section~2 describes our numerical methods.
Results are given in section~3.  
 
\section{Numerics and Initial Conditions}
\label{sec:num}

We use the $N$-body part of the tree-particle-mesh Smoothed Particle Hydrodynamics code 
GADGET-3 originally described in Springel (2005). The units of mass and distance are taken as 
$10^{11}\,M_\odot$ and 1\,kpc, respectively. 
We use $N_{\rm h} = 10^6$ particles for the DM halo, and  $N_{\rm d} = 2\times 10^5$ for stars.
Convergence models have been run with $N_{\rm h} = 4\times 10^6$ and $N_{\rm d} = 4\times 10^5$,
in compliance with the Dubinski et al. (2009) study of discrete resonance interactions
between the bar and halo orbits.
The gravitational softening is $\epsilon_{\rm grav}=50$\,pc for stars and DM.
To simplify the analysis we have ignored the stellar bulge. The opening angle $\theta$ of the
tree code has been reduced from 0.5 used in cosmological simulations to 0.4 which increases
the quality of the force calculations. Our models
have been run for 10\,Gyr with an energy conservation of 0.08\% and angular momentum
conservation of 0.05\% over this time. 

To construct the initial conditions, we have used a novel method introduced by Rodionov \& Sotnikova
(2006), see also Rodionov et al. (2009). We provide only minimal details for this method,
which is elaborated elsewhere. It is based on the constrained
evolution of a dynamical system. The basic steps include (1) constructing the model using prescribed
positions of the particles with some (non-equilibrium) velocities, (2) allowing the particles to evolve 
for a short time which leads to modified positions and velocities, (3) returning the particles
to the old positions with the new velocities, and (4) iterating on the previous steps until
velocities converge to equilibrium values. This results in the near-equilibrium dynamical system
which is then evolved.  

The initial disk has been constructed as exponential, with the volume density given by 

\begin{equation}
\rho_{\rm d}(R,z) = \biggl(\frac{M_{\rm d}}{4\pi h^2 z_0}\biggr)\,{\rm exp}(-R/h) 
     \,{\rm sech}^2\biggl(\frac{z}{z_0}\biggr),
\end{equation}
where $M_{\rm d}=6.3\times 10^{10}\,M_\odot$ is the disk mass, $h=2.85$\,kpc is its radial 
scalelength, and $z_0=0.6$\,kpc is the scaleheight. $R$ and $z$ represent the cylindrical coordinates. 

The halo density is given by Navarro, Frenk \& White (1996, NFW):

\begin{equation}
\rho_{\rm h}(r) = \frac{\rho_{\rm s}\,e^{-(r/r_{\rm t})^2}}{[(r+r_{\rm c})/r_{\rm s}](1+r/r_{\rm s})^2}
\end{equation}
where $\rho(r)$ is the DM density in spherical coordinates, $\rho_{\rm s}$
is the (fitting) density parameter, and $r_{\rm s}=9$\,kpc is the characteristic radius, where the power 
law slope is (approximately) equal
to $-2$, and $r_{\rm c}$ is a central density core. We used the Gaussian cutoffs at 
$r_{\rm t}=86$\,kpc for the halo and $R_{\rm t}=6h\sim 17$\,kpc
for the disk models, respectively. The halo mass is $M_{\rm h} = 6.3\times 10^{11}\,M_\odot$,
and halo-to-disk mass ratio within $R_{\rm t}$ is $\sim 2$.  Other ratios have
been explored as well. Oblate halos with various 
polar-to-equatorial axis ratios, $q=c/a$, 
have been analyzed, with $0.8\ltorder q\ltorder 1$. Here, we limit our discussion to cuspy halos 
with $q\sim 1$, and a small 
core of $r_{\rm c}=1.4$\,kpc. Other profiles, such as the large core NFW and
isothermal sphere density profiles, have been implemented as well, and
resulted in qualitatively similar evolution. Dispersion velocity anisotropy, $\beta$, has been 
constrained initially to be mild, using the novel method of Constrained Evolution 
discussed above. Velocities have been taken to be isotropic in the central region and the
anisotropy increased to $\beta\sim 0.3$ outside the disk.

Disk radial dispersion velocities have been taken as $\sigma_{\rm R}(R)= \sigma_{\rm R,0}\,{\rm
exp}(-R/2h)$ with  $\sigma_{\rm R,0}=143\,{\rm km\,s^{-1}}$. This results in $Q=1.5$
at $R\sim 2.42\,h$, and increasing values toward the center and outer disk. Vertical velocity
dispersions are $\sigma_{\rm z}(R)=\sigma_{\rm z,0}\,{\rm exp}(-R/2h)$, with  $\sigma_{\rm
z,0}=98\,{\rm km\,s^{-1}}$.

To form spinning halos, we have flipped the angular momenta, $J_{\rm z}$, 
of a prescribed fraction of DM particles which are on retrograde orbits with respect to the disk, 
by reversing their velocities, in line with Lynden-Bell's (1960) Maxwell demon. Only $\lambda\sim 0-0.09$ 
models are discussed here. 
The $\lambda < 0$ cases are simpler, due to a decreased fraction of prograde halo particles able to
resonate with the bar/disk particles (e.g., Christodoulou et al. 1995). 
The implemented velocity reversals preserve the solution to the Boltzmann
equation and do not alter the DM density profile or velocity magnitudes (e.g., Lynden-Bell 1960, 1962; 
Weinberg 1985). For spherical halos, the invariancy under velocity reversals is a direct corollary of 
the Jeans (1919) theorem (see also Binney \& Tremaine 2008). The most general distribution function 
for such systems is a sum  of $f(E,J^2)$, where $E$ is the energy, $J$ --- the value of the total angular 
momentum (i.e., $J^2$), and of an odd function of $J_{\rm z}$, 
i.e.,  $g(E,J,J_{\rm z})$ (Lynden-Bell 1960). If $g\neq 0$, the spherical system has 
a net rotation around this axis.

We left the disk parameters unchanged, while halo models have varied spin
$\lambda$. The value of $\lambda$ has been added to the model name using the last two significant 
digits, e.g., P60 means $\lambda=0.060$ and ``P'' stands for prograde.   

\section{Results}
\label{sec:res}

\begin{figure}
\begin{center}
\includegraphics[angle=0,scale=0.47]{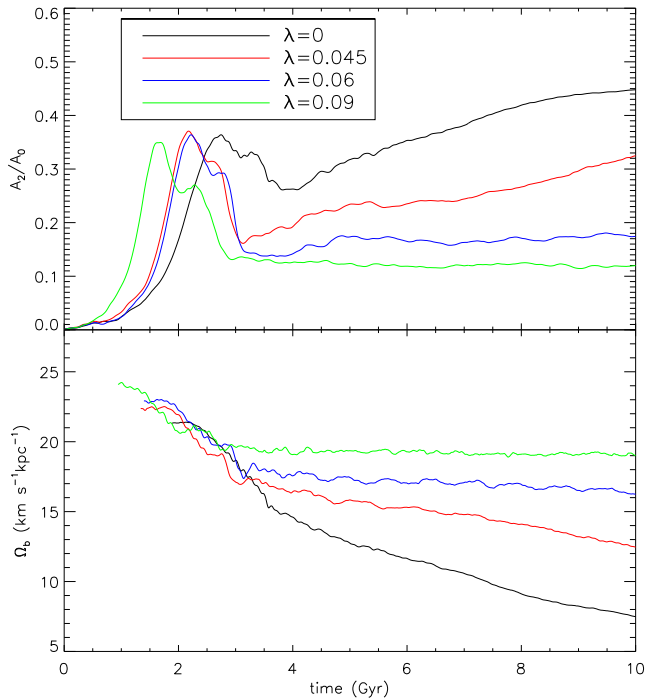}
\end{center}
\caption{{\it Upper:} Evolution of the bar amplitudes, $A_2$ (normalized by the monopole term $A_0$), for  
for spherical NFW  halos with $q=1$.
Shown are P00, P45, P60 and P90 models. {\it Lower:} Evolution of bar pattern speed, $\Omega_{\rm b}$, 
in the above models.
}
\end{figure}

\begin{figure*}
\begin{center}
\includegraphics[angle=0,scale=0.9]{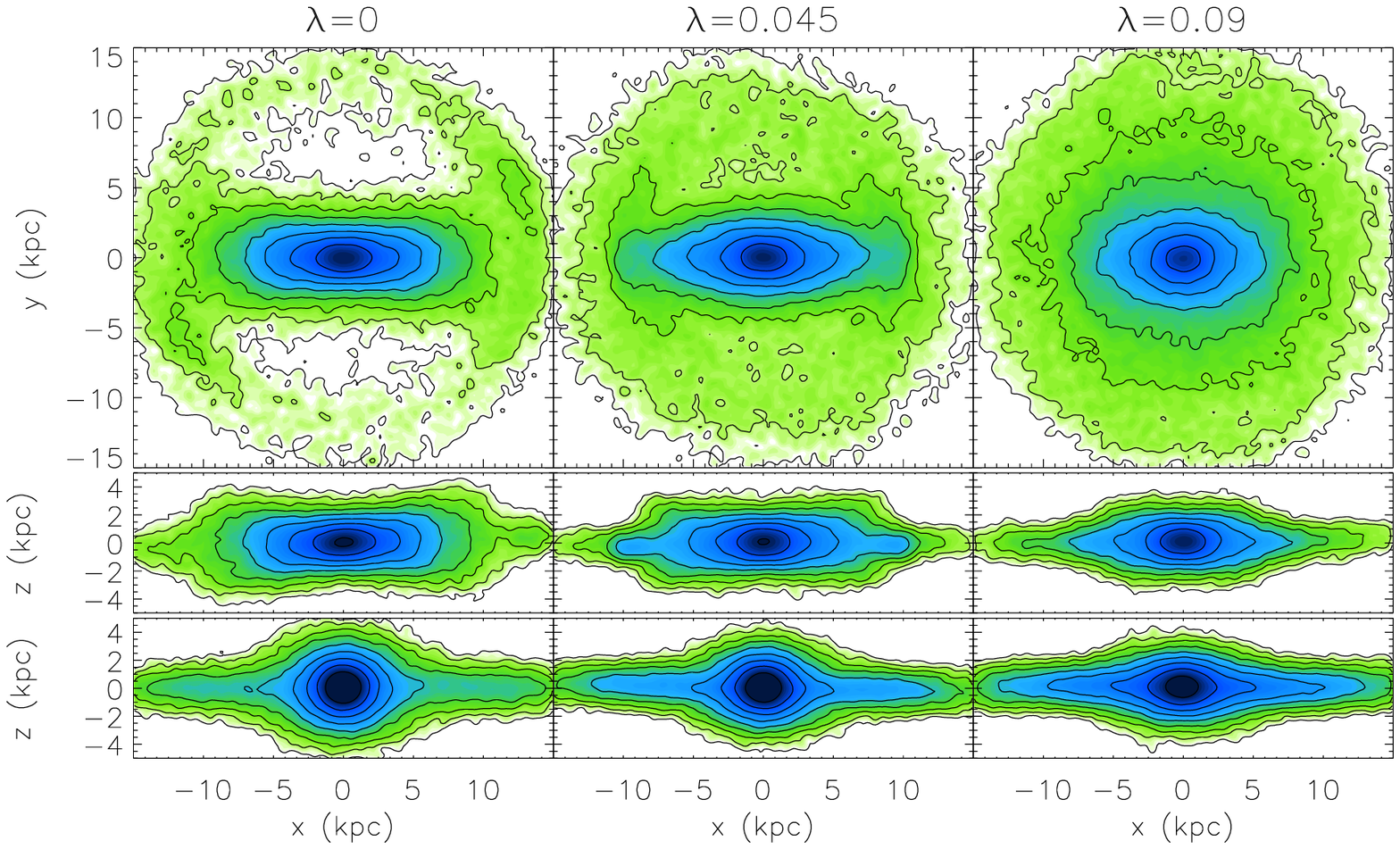}
\end{center}
\caption{Disk-bar surface density contours (face-on, edge-on, and end-on) at $t=10$\,Gyr, for the NFW 
halos with $q=1$, P00 (left column), P45 (center) and P90 (right) models. Note the different bulge
shapes: X-shaped for P00, boxy/X-shaped for P45, and boxy for P90, as well as decreasing strength of
{\it ansae} with increasing $\lambda$ (see text).
}
\end{figure*}

All models presented here have an identical
mass distribution, both in DM and stars. Hence, any differences in the
evolution must follow from the initial distribution of angular momentum in DM halos and
its redistribution in the bar-disk-halo system.
Figure\,1 displays the evolution of the stellar bars through the amplitudes of the Fourier $m=2$ mode, 
$A_2$, 
and their pattern speeds, $\Omega_{\rm b}$, for 10\,Gyrs. This timescale is probably close to the maximum 
uninterrupted growth of galactic disks in the cosmological framework, and hence to the lifetime of
the bars. The normalized (by the monopole term $A_0$) bar amplitude has been defined here as 

\begin{equation}
\frac{A_2}{A_0} = \frac{1}{A_0}\sum_{i=1}^{N_{\rm d}} m_{\rm i}\,e^{2i\phi_{\rm i}},
\end{equation} 
for $R\leq 14$\,kpc. The summation is performed over all disk particles with the mass $m=m_{\rm i}$
at angles $\phi_{\rm i}$ in the disk plane. $\Omega_{\rm b}$ is obtained
from the phase angle $\phi= 0.5\,{\rm tan^{-1}}[{\rm Im}(A_2)/{\rm Re}(A_2)]$ evolution with time.
We divide the evolution into two phases: the dynamical phase, which consists of the initial 
bar instability and terminates with the vertical buckling instability of the bars and formation of 
boxy/peanut-shaped bulges (e.g., Combes et al. 1990; Pfenniger \& Friedli 1991; Raha et al. 1991; 
Patsis et al. 2002; 
Athanassoula 2005; Berentzen et al. 2007). Buckling weakens the bar but does not dissolve it 
(Martinez-Valpuesta \& Shlosman 2004). Repeated bucklings increase the size of the bulge 
(Martinez-Valpuesta \&  Shlosman 2005; Martinez-Valpuesta et al. 2006). One buckling has been
observed in the models presented here --- following it,
the bar enters the second phase, that of secular evolution.

The most striking development observed in models of Figure\,1 during the secular phase is an increased 
damping of the bar amplitude and a slower or absent bar growth for $\lambda\gtorder 0.03$. 
The P00 model ($\lambda=0$) displays healthy growth after buckling.
The P30 and P45 bars have a slower growth rate than the P00 bar, and do not recover their pre-buckling 
strength even after 10\,Gyr.  But models P60 and P90 show no growth in $A_2$ at all. 
The corresponding pattern
speed evolution, $\Omega_{\rm b}(t)$, for these models differs substantially as well. The A90 bar 
displays a perfectly flat $\Omega_{\rm b}(t)$, and does not lose its angular momentum to the
disk and/or the halo. This includes both the internal angular momentum (i.e., circulation) and the
tumbling. Similar trend between the final $\Omega_{\rm b}$ and $\lambda$ can be also
observed in Figure\,7 of Debattista \& Sellwood (2000), although low
resolution apparently prevented any conclusion of this sort.
 
Figure\,2 compares the end products of the secular evolution of barred disks in models P00, P45 and P90. 
The differences appear to be profound. First, the bar size clearly anticorrelates with $\lambda$ --- 
this is
a reflection of the inability of the bar potential to capture additional orbits and grow in length and
mass. Second, the {\it ansae} (handles) feature is the strongest in the P00 bar, while it is smaller in 
size for P45 
and completely absent in the P90 bar. Ansae have been associated with captured disk orbits librating around
the bar (Martinez-Valpuesta 2006; Martinez-Valpuesta et al. 2006). This is another indication that
the bar in high-$\lambda$ models does not grow. Note that the surface density in the disk is clearly affected,
as trapping of the disk orbits by the P00 bar creates low-density regions in the disk but not in P90. We
analyzed the properties of the halo `ghost' bar (Holley-Bockelmann et al. 2005; Athanassoula
2007; Shlosman 2008), and found no growth there as well. The offset angle between the ghost and stellar
bars remains near zero (within the error margin). Third, the face-on morphology of the P00 bar is that of a 
rectangular shape, while that of P90 is elliptical. Fourth, bulges that formed as a a result of the buckling
instability show the same anticorrelation trend in size$-\lambda$, as seen in edge-on (i.e., along the bar's 
minor axis) frames. Furthermore, they differ 
in shape as well: the P00 bulge has an X-shape, P45 is boxy/X-shaped, and P90 is boxy. Trapped 3-D orbits
are responsible for the bulge shape (e.g., Patsis et al. 2002; Athanassoula 2005; Martinez-Valpuesta et al. 
2006). 

What is even more intriguing is the near or complete absence of secular braking in the P60 and P90 bars. Although the
bars are weak, constancy of $\Omega_{\rm b}$ and $A_2$ over 6\,Gyr in P90 points to no angular momentum 
transfer away from the bar, or, alternatively, to an opposite flux from the halo which compensates for the 
loss of angular momentum by the bar. As we see below, it is the second possibility that takes place.
While the P60 and P90 models exhibit extremes of this effect, it is visible at various levels in all models 
with $\lambda\gtorder 0.02$. 

While most of the angular momentum transfer away from the bar is due to resonances, we 
deal with this aspect of the problem elsewhere. However, we do quantify the {\it rate} of the
overall angular momentum transfer between the disk and the halo, i.e., accounting for the resonant and
non-resonant angular momentum redistribution. This is accomplished by dividing the disk and halo into
nested cylindrical shells and constructing a two-dimensional map of the angular momentum change in each shell as
a function of $R$ and $t$ (e.g., Villa-Vargas et al. 2009, 2010). Such a color-coded diagram is shown in 
Figure\,3 for disk stars (lower frames), $\langle\dot J_*\rangle\equiv (\partial J_*/\partial t)_{\rm R}$ 
and for halo particles (upper frames), 
$\langle\dot J_{\rm DM}\rangle\equiv (\partial J_{\rm DM}/\partial t)_{\rm R}$, 
where the brackets indicate time-averaging.  

\begin{figure*}
\begin{center}
\includegraphics[angle=-90,scale=0.6]{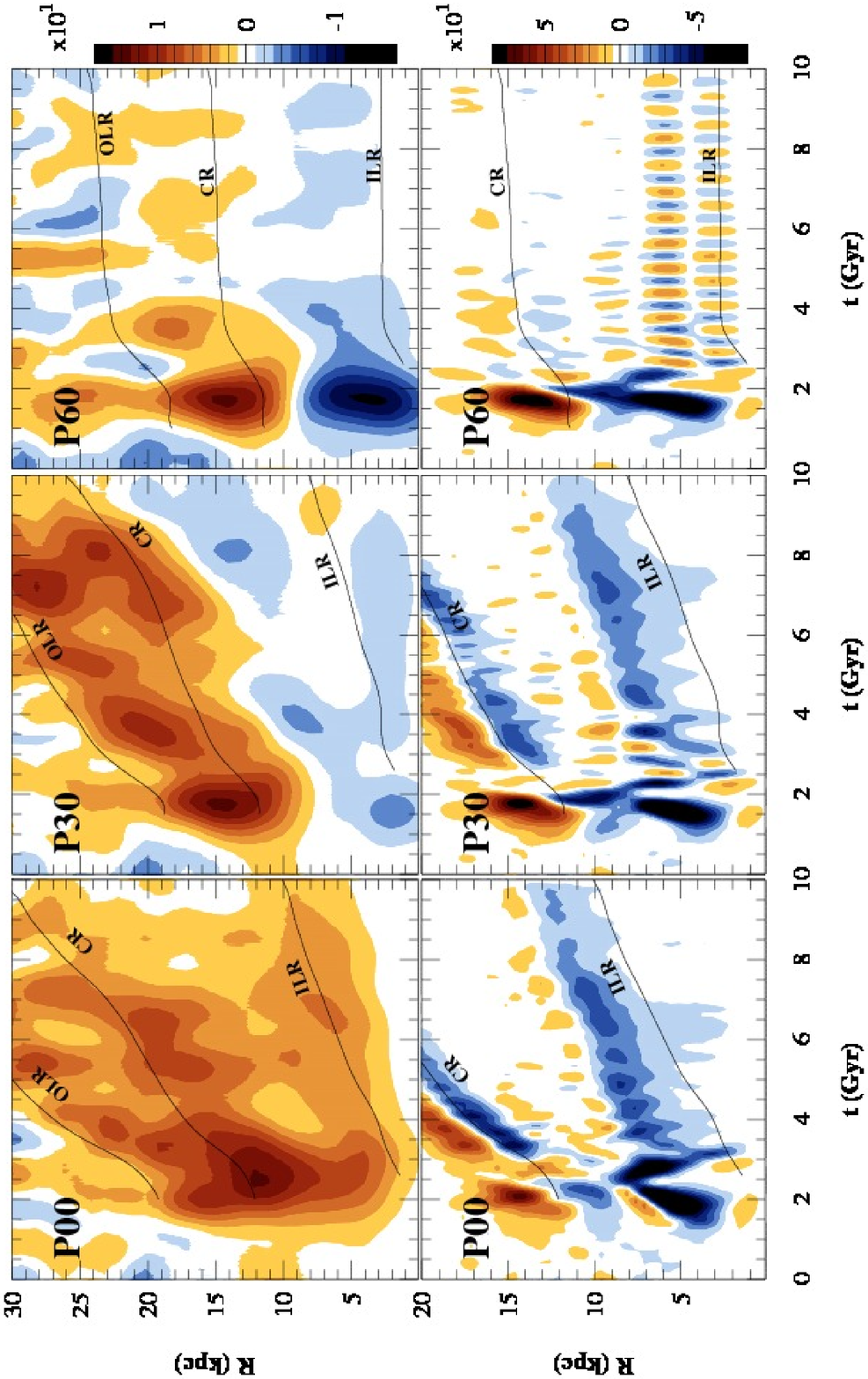}
\end{center}
\caption{\underline{DM halos} ({\it upper frames}): Rate of angular momentum flow $\dot J$ as a function of 
a cylindrical 
radius and time for the P00 (left), P30 (middle), and P60 (right) models with $q=1$ NFW DM halos. 
The color palette corresponds to gain/loss rates (i.e., red/blue) using a logarithmic scale in color. 
The cylindrical shells have $\Delta R = 1$\,kpc, extending to $z=\pm \infty$. 
\underline{Stellar disks} ({\it lower frames}): same for (identical) disk models embedded in the P00 (left), 
P30 (middle), and 
P60 (right) halos, except $\Delta R = 0.5$\,kpc, and $|\Delta z| = 3$\,kpc. Positions of major disk 
resonances, ILR, CR, and OLR, have been marked.
}
\end{figure*}

The diagrams for P00 are the easiest to understand. The red (blue) colors correspond to the absorption
(emission) of the angular momentum. The continuity of these colors for the P00 disk represents the emission
and absorption of angular momentum by the disk prime resonances. For example, the dominant blue band drifting 
to larger
$R$ with time is associated with the emission of angular momentum by the inner Lindblad resonance (ILR),
and the additional blue band corresponds to the Ultra-Harmonic resonance (UHR). The dominant red band
follows the corotation resonance (CR) and the outer Lindblad resonance (OLR). 

The number of DM particles on prograde orbits has steadily increased with $\lambda$, raising the
possibility of resonant coupling between them and the bar orbits. This is supported by linear theory
(Weinberg 1985 and refs. therein) and by numerical simulations (Saha \& Naab 2013). 
Indeed, we observe increased emission of angular momentum by the ILR and corresponding enhanced absorption
by the halo. Halo particles are late to pick up the angular momentum
from the bar (due to their higher velocity dispersion), but the exchange is visible already before buckling. 
Enhanced coupling between the orbits is the reason for the shorter timescale for bar instability.

The secular evolution of bars, however, proceeds under quite different conditions. The bar cannot be 
considered as a linear perturbation, and the halo orbits have already been heavily perturbed and some have 
been captured by the stellar bar. So, one expects the nearby halo orbits around the bar to be tightly
correlated with the bar. The upper frames in Figure\,3 display the rate of angular momentum flow in the DM halo.
While the P00 halo appears to be completely dominated by the absorption of angular momentum at all major
resonances (ILR, CR, OLR), P30 shows a quite different behavior and emits it at the ILR. The loss
of angular momentum in this region of the DM halo is even more intense in P90. 
Already at the buckling, we can observe a weak blue band of emission in the P30 halo, alongside a strong
absorption, instead of pure absorption in P00. Note that linear resonances shown by continuous curves
appear to be a bad approximation to the actual nonlinear resonances given by the color bands,
because they are calculated under assumption of circular orbits. In the P90 halo, a strong {\it emission} 
is visible at the position of the disk ILR, which continues as a band  
alongside weakened absorption. So the absorption gradually weakens and moves out with increased $\lambda$, 
while the emission strengthens and spreads. The disk emission and absorption by major resonances 
also differs with changing $\lambda$ --- it gradually develops an intermittent behavior, especially
at the ILR in P60, where the blue and red bands become intermittent. Such a cyclical behavior is not seen 
in the P00 disk, but becomes visible in the P30 disk and dominates the inner P60 disk. Hence, the spinning halo 
appears to emit and absorb angular momentum recurrently. The halo as a whole still absorbs the
angular momentum from the disk in P30, while the net flux is zero for P90.

This result is anticipated. The ability to pump angular momentum into a selected
number of halo particles by means of a stellar bar is not without limits. As the angular momentum of the
prograde population in the halo is increased, its ability to absorb angular momentum should saturate, and,
under certain conditions, even be reversed. After buckling, the bar weakens substantially, as seen in
Figure\,1. At later stages, as the bar is expected to resume its growth, the near (disk) halo orbits can possess
more angular momentum than the bar region which has been losing it for some time. For this prograde 
population, increase of $\lambda$ simply
increases the initial angular momentum and the saturation comes earlier. {\it What emerges as a fundamental
property of a DM halo is the angular momentum and its distribution for the prograde population of 
orbits,  irrespective of the value of $\lambda$.}

Evolution of galactic bars is inseparable from the cosmological evolution of their host galaxies. We find
that the secular growth of bars is significantly anticorrelated with the halo spin for $\lambda\gtorder
0.03$. This means that majority of halos will adversely affect the bar strength, and, therefore, the
angular momentum transfer and the bar braking. 
Beyond dynamical consequences, bars in spinning halos will be systematically smaller, which 
will make their detection at larger redshifts more difficult. This trend can be further strengthened 
because, during mergers, for a limited time period of $\sim 1-2$\,Gyr, $\lambda$ has been shown to
increase (e.g., Hetznecker \& Burkert 2006; Romano-Diaz et al. 2007; Shlosman 2013). Weaker bars
are known to possess star formation along the offset shocks, unlike strong bars, and are less 
efficient in moving the gas inward. Furthermore, damping bar amplitude has implications for disk
morphology, stellar populations and abundance gradients. 

To summarize, we have investigated the dynamical and secular evolution of stellar bars in spinning DM halos. 
In a representative set of numerical models, we find that
the angular momentum flow in the disk-halo system is substantially affected by the momentum distribution
in the prograde population of DM particles, and is not limited
to the momentum flux from the disk to halo. The associated bar pattern speed slowdown is minimized
and ceases for larger $\lambda$. This means that the bar does not experience gravitational torques
and its amplitude remains steady, while the angular momentum, both internal circulation and tumbling, is 
preserved. This trend becomes
visible for $\lambda\gtorder 0.02$ and dominates the bar evolution for halos with $\lambda\gtorder 0.03$.
Because of a lognormal distribution of $\lambda$ with a mean value of $0.035\pm 0.005$, a substantial 
fraction of DM halos will be affected. We analyze the
rate of angular momentum change by subdividing the disk-halo system into nested cylindrical shells, and 
show that the DM {\it halo can both absorb and emit angular momentum}, resulting in a reduction of 
the net transfer of angular momentum from the disk to the halo.
The ability of the halo material to both emit and absorb angular momentum has important corollaries.

\acknowledgements 
We are grateful to Sergey Rodionov for guidance with the iterative method to construct initial 
conditions, and to Ingo Berentzen, Jun-Hwan Choi, Emilio Romano-Diaz, Raphael Sadoun and Jorge 
Villa-Vargas for help with numerical
issues. We thank Volker Springel for providing us with the original version of GADGET-3. This work 
has been partially supported by grants from the NSF and the STScI (to I.S.). Simulations have been
performed on the University of Kentucky DLX Cluster.


\begin{thebibliography}{}
\expandafter\ifx\csname natexlab\endcsname\relax\def\natexlab#1{#1}\fi 

\bibitem[{{Athanassoula} (2002)}]{Athanassoula:02}
{Athanassoula}, E., 2002, \apjl, 569, L83

\bibitem[{{Athanassoula} (2003)}]{Athanassoula:03}
{Athanassoula}, E., 2003, \mnras, 341, 1179

\bibitem[{{Athanassoula} (2005)}]{Athanassoula:05}
{Athanassoula}, E., 2005, \mnras, 358, 1477

\bibitem[{{Athanassoula} (2007)}]{Athanassoula:07}
{Athanassoula}, E., 2007, \mnras, 377, 1569

\bibitem[{{Athanassoula} {et~al.}(2013)}]{Athanassoula.etal:13}
{Athanassoula}, E., {Machado}, R.E.G., \& {Rodionov}, S.A. 2013, \mnras, 429, 1949

\bibitem[{{Barnes} \& {Efstathiou}(1987)}]{Barnes.Efstathiou:87}
{Barnes}, J., \& {Efstathiou}, G. 1987, \apj, 319, 575

\bibitem[{{Berentzen} \& {Shlosman}(2006)}]{Berentzen.Shlosman:06}
{Berentzen}, I., \& {Shlosman}, I. 2006, \apj, 648, 807 

\bibitem[{{Berentzen} {et~al.}(2006)}]{Berentzen.etal:06}
{Berentzen}, I., {Shlosman}, I. \& {Jogee}, S. 2006, \apj, 637, 582
 
\bibitem[{{Berentzen} {et~al.}(2007)}]{Berentzen.etal:07}
{Berentzen}, I., {Shlosman}, I., {Martinez-Valpuesta}, I., \& {Heller}, C. 2007, \apj, 666, 189

\bibitem[{{Binney} \& {Tremaine}(2008)}]{Binney.Tremaine:08}
{Binney}, J., \& {Tremaine}, S. 2008, Galactic Dynamics (Princeton, NJ: Princeton Univ. Press)

\bibitem[{{Bullock} {et~al.}(2001)}]{Bullock.etal:01}
{Bullock}, J.S., {Dekel}, A., {Kolatt}, T.S., {Kravtsov}, A.V., {Klypin}, A.A., {Porciani}, C., 
   {Primack}, J.R. 2001, \apj, 555, 240

\bibitem[{{Christodoulou} {et~al.}(1995)}]{Christodoulou.etal:95}
{Christodoulou}, D.M., {Shlosman}, I., \& {Tohline}, J.~E. 1995, \apj, 443, 551

\bibitem[{{Combes} {et~al.}(1990)}]{Combes.etal:90}
{Combes}, F., {Debbash}, F., {Friedli}, D., \& {Pfenniger}, D. 1990, A\&A, 233, 82

\bibitem[{{Debattista} \& {Sellwood}(1998)}]{Debattista.Sellwood:98}
{Debattista} V., \& {Sellwood}, J.A. 1998, \apj, 493, 5

\bibitem[{{Debattista} \& {Sellwood}(2000)}]{Debattista.Sellwood:00}
{Debattista} V., \& {Sellwood}, J.A. 2000, \apj, 543, 704

\bibitem[{{Dubinski} {et~al.}(2009)}]{Dubinski.etal:09}
{Dubinski}, J., {Berentzen}, I., \& {Shlosman}, I. 2009, \apj, 697, 293

\bibitem[{{El-Zant} \& {Shlosman}(2002)}]{El-Zant.Shlosman:02}
{El-Zant}, A., \& {Shlosman},I. 2002, \apj, 577, 626

\bibitem[{{El-Zant} {et~al.}(2003)}]{El-Zant.etal:03}
{El-Zant}, A., {Shlosman}, I., {Begelman}, M.C., \& {Frank}, J. 2003, \apj, 590, 641

\bibitem[{{Heller} {et~al.}(2007)}]{Heller.etal:07}
{Heller}, C.H., {Shlosman}, I., \& {Athanassoula}, E. 2007, \apj, 671, 226

\bibitem[{{Hetznecker} \& {Burkert}(2006)}]{Hetznecker.Burkert:06}
{Hetznecker} H., \& {Burkert}, A. 2006, \mnras, 370, 1905

\bibitem[{{Holley-Bockelmann} {et~al.}(2005)}]{Holley.etal:05}
{Holley-Bockelmann}, K., {Weinberg}, M.D, \& {Katz}, N. 2005, \mnras, 363, 991

\bibitem[{{Hoyle} (1949)}]{Hoyle:49} 
{Hoyle}, F. 1949, in Problems in Cosmical Aerodynamics, ed. J.M. Burgers \& H.C. van de
Hulst (Dayton, OH: Central Air Documents Office, Dayton, OH), p.\,19

\bibitem[{{Jeans}(1919)}]{Jeans:19}
{Jeans}, J.H. 1919, Problems of Cosmogony and Stellar Dynamics (Cambridge: Cambridge
Univ. Press)

\bibitem[{{Lynden-Bell} (1960)}]{Lynden-Bell:60}
{Lynden-Bell}, D. 1960, \mnras, 120, 204

\bibitem[{{Lynden-Bell} (1962)}]{Lynden-Bell:62}
{Lynden-Bell}, D. 1962, \mnras, 123, 447

\bibitem[{{Lynden-Bell} \& {Kalnajs}(1972)}]{Lynden-Bel.Kalnajs:72}
{Lynden-Bell}, D., \& {Kalnajs}, A.~J. 1972, \mnras, 157, 1

\bibitem[{{Machado} \& {Athanassoula}(2010)}]{Machado.Athanassoula:10}
{Machado}, R.E.G., \& {Athanassoula}, E 2010, \mnras, 406, 2386 

\bibitem[{{Martinez-Valpuesta} (2006)}]{Martinez-Valpuesta:06}
{Martinez-Valpuesta}, I. 2006, Ph.D. Thesis, University of Hertfordshire

\bibitem[{{Martinez-Valpuesta} \& {Shlosman}(2004)}]{Martinez-Valpuesta.Shlosman:04}
{Martinez-Valpuesta}, I., \& {Shlosman}, I. 2004, \apjl, 613, L105 

\bibitem[{{Martinez-Valpuesta} \& {Shlosman}(2005)}]{Martinez-Valpuesta.Shlosman:05}
{Martinez-Valpuesta}, I., \& {Shlosman}, I. 2005, in AIP Conf. Proc. 783, The Evolution of 
    Starbursts, ed. S. H\"uttemeister et al.,  p.\,189

\bibitem[{{Martinez-Valpuesta} {et~al.}(2006)}]{Martinez-Valpuesta.etal:06}
{Martinez-Valpuesta}, I., {Shlosman}, I., \& {Heller}, C. 2006, \apj, 637, 214

\bibitem[{{Navarro} {et~al.}(1996)}]{Navarro.etal:96}
{Navarro}, J.F., {Frenk}, C.S., \& {White}, S.D.M. 1996, \apj, 462, 563 (NFW)

\bibitem[{{Patsis} {et~al.}(2002)}]{Patsis.etal:02}
{Patsis}, P.A., {Skokos}, Ch., \& {Athanassoula}, E. 2002, \mnras, 337, 578

\bibitem[{{Pfenniger} \& {Friedli}(1991)}]{Pfenniger.Friedli:91}
{Pfenniger}, D., \& {Friedli}, D. 1991, A\&A, 252, 75

\bibitem[{{Porciani} {et al.}(2002)}]{Porciani.etal:02}
{Porciani}, C., {Dekel}, A., \& {Hoffman}, Y. 2002, \mnras, 332, 325

\bibitem[{{Raha} {et~al.}(1991)}]{Raha.etal:91}
{Raha}, N., {Sellwood}, J.A., {James}, R.A., \& {Kahn}, F.D. 1991, Nature, 352, 411

\bibitem[{{Romano-Diaz} {et al.}(2007)}]{Romano-Diaz.etal:07}
{Romano-Diaz}, E., {Hoffman}, Y., {Heller}, C., {Faltenbacher}, A., {Jones}, D.
     \& {Shlosman}, I. 2007, \apj, 657, 56

\bibitem[{{Rodionov} {et~al.}(2009)}]{Rodionov.etal:09}
{Rodionov}, S.A., {Athanassoula}, E., \& {Sotnikova}, N.Ya. 2009, \mnras, 392, 904

\bibitem[{{Rodionov} \& {Sotnikova}(2006)}]{Rodionov.Sotnikova:06}
{Rodionov}, S.A., \& {Sotnikova}, N.Ya. 2006, ARep, 50, 893

\bibitem[{{Saha} \& {Naab}(2013)}]{Saha.Naab:13}
{Saha}, K., \& {Naab}, T. 2013, \mnras, 434, 1287

\bibitem[{{Sellwood}(1980)}]{Sellwood:80}
{Sellwood}, J.A. 1980, A\&A, 89, 296

\bibitem[{{Shlosman}(2008)}]{Shlosman:08}
{Shlosman}, I. 2008, in ASP Conf. Ser. 390, Pathways Through an Eclectic Universe, ed. J.H.Knapen et al.,
        p.\,440, arXiv:0710.0630

\bibitem[{{Shlosman}(2013)}]{Shlosman:13}
{Shlosman}, I. 2013, in Secular Evolution of Galaxies, ed. J.Falcon-Barroso \& J.H.Knapen, 
        (Cambridge: Cambridge Univ. Press), p.\,555, arXiv:1212.1463

\bibitem[{{Springel}(2005)}]{Springel:05} 
{Springel}, V. 2005, \mnras, 364, 1101

\bibitem[{{Tremaine} \& {Weinberg}(1984)}]{Tremaine.Weinberg:84}
{Tremaine}, S., \& {Weinberg}, M.D. 1984, \mnras, 209, 729 

\bibitem[{{Tremaine} \& {Ostriker}(1999)}]{Tremaine.Ostriker:99}
{Tremaine}, S., \& {Ostriker}, J.P. 1999, \mnras, 306, 662

\bibitem[{{Villa-Vargas} {et~al.}(2009)}]{Villa-Vargas.etal:09}
{Villa-Vargas}, J., {Shlosman}, I., \& {Heller}, C.H. 2009, \apj, 707, 218

\bibitem[{{Villa-Vargas} {et~al.}(2010)}]{Villa-Vargas.etal:10}
{Villa-Vargas}, J., {Shlosman}, I., \& {Heller}, C.H. 2010, \apj, 719, 1470

\bibitem[{{Weinberg}(1985)}]{Weinberg:85}
{Weinberg}, M.D. 1985, MNRAS, 213, 451

\bibitem[{{Weinberg} \& {Katz}(2007)}]{Weinberg.Katz:07}
{Weinberg}, M.D., \& {Katz}, N.. 2007, \mnras, 375, 460 

\bibitem[{{White}(1978)}]{White:78}
{White}, S.D.M. 1978, \mnras, 184, 185


\end{thebibliography}
\end{document}